\documentclass[twocolumn,letter]{jpsj3}
\usepackage{txfonts}
\usepackage{color}
\usepackage{amssymb}
\usepackage{amsmath}

\title{Efficiency of Free Energy Transduction in Autonomous Systems}

\author{\name{Kyogo \surname{Kawaguchi}}\thanks{E-mail address: kyogok@daisy.phys.s.u-tokyo.ac.jp} and \name{Masaki \surname{Sano}}\thanks{E-mail address: sano@phys.s.u-tokyo.ac.jp}
}

\inst{Department of Physics, The University of Tokyo, \address{Hongo 7-3-1, Tokyo 113-0033, Japan}}
\abst{
We consider the thermodynamics of chemical coupling from the viewpoint of free energy transduction efficiency. In contrast to an external parameter-driven stochastic energetics setup, the dynamic change of the equilibrium distribution induced by chemical coupling, adopted, for example,
in biological systems, is inevitably an autonomous process.
We found that the efficiency is bounded by the ratio between the non-symmetric and the symmetrized Kullback-Leibler distance, which is significantly lower than unity.
Consequences of this low efficiency are demonstrated in the simple two-state case, which serves as an important minimal model for studying the energetics of biomolecules.}

\kword{chemical coupling, efficiency bound, stochastic thermodynamics, biophysics, information theory}

\begin{document}
\maketitle

\textit{Introduction.}---Biological phenomena are based on nano-micron-scaled mechanics and reactions. Physics of the fundamental processes occuring inside cells, such as  transcription/translation dynamics or motions of molecular motors~\cite{howard} and ion pumps~\cite{astumian1,astumian2}, should all be based on small-system's thermodynamics. Recent developments in the study of thermodynamics on this scale, including non-equilibrium work relations~\cite{jar,trep,crooks,hatano}, treatments of feedback control~\cite{dem,demex}, and generalized fluctuation-dissipation relations~\cite{harasasa,gomez,F1}, are well supported by experimental techniques at the single-molecule level. However, the direct relationship between the stochastic thermodynamics scheme~\cite{bustamante,stoch,sekimoto} and the actual biological reactions is unclear, and the physical design principles of biomolecular dynamics remain a challenge. 

A large gap exists in the treatment of external parameter between the knowledge of stochastic thermodynamics and actual biophysical reactions.
In a typical setting, such as cases of the Jarzynski equation~\cite{jar} and the fluctuation theorem~\cite{crooks}, 
a set procedure of externally switching the parameter commonly written as $\lambda$ is considered.
Such non-fluctuating external control is experimentally achieved by the intervention of a macroscopic setup, as in the case of laser traps and magnetic or electric fields acting on small beads~\cite{trep,demex,gomez}.
On the other hand, biomolecular dynamics is typically autonomous with only limited ability to control the probabilities of desired and undesired events. 
In an autonomous and substantially isothermal system, the source of this operation must suffer the same order of thermal fluctuations as the target, thus forcing biological systems to adopt a chemical coupling strategy~\cite{gaspard1,schmiedl}.

Out-of-equilibrium free energy is the source of chemical-coupling-driven control.
Altering  equilibrium situations in a successive manner for producing irreversible dynamics 
shall be physically viewed as a free energy transduction phenomena.
This is significant in that delivering free energy by this way is different to simply treating chemical energy as a source of force and thermodynamic work.
Specifically, transduction efficiency in such a scheme may not, in principle, reach the second law limit as is the case with the proposed operational setup.

In this letter, we discuss the thermodynamics of such autonomous free energy transduction.
We consider the general case of altering an equilibrium distribution for achieving a new equilibrium situation through a chemical coupling procedure,
and derive a bound on its efficiency (Eq.~(\ref{eq:eff})).
Resulting expressions quantify the intuition of the efficiency falling significantly below the second law bound.
The substantial reduction in efficiency is demonstrated for a specific case of the two-state model.

\textit{Efficiency of free energy transduction.}---
We set up a system that undergoes Markov transitions between discrete energy (or constrained free energy)
states. $i, (=1, \cdots, N)$. We consider linking this system to a chemical potential reservoir (i.e., the outer setup) and transducing free energy into it by altering its detailed balance condition.
Initially, the system is isolated from the chemical potential bath, characterized by the probability distribution
$\{ p_i \}$ ($p_i= e^{-\beta U_i}/Z$), over the states with energy $U_i$ in thermal equilibrium at constant temperature $\beta^{-1} (Z= \sum_{i} e^{-\beta U_i})$.
For reaching the new distribution, $\{ q_i \}$, the system is then placed in contact with the chemical bath in which the transition rates are associated with coupled chemical reactions.
Comparing the free energy gain in the system, $\Delta F$,
with the free energy consumed in the outer setup (chemical baths), $\Delta F_c$, we 
determine that the efficiency of free energy transduction $\sigma_f$ is bounded by an inequality,
\begin{equation}
\sigma_f \equiv \frac{\Delta F}{\Delta F_c} \leq \frac{D(q||p)}{D(q||p)+D(p||q)} \leq 1. \label{eq:eff}
\end{equation}
Here $D(q||p)$ is the relative entropy, or the Kullback-Leibler distance between the distributions  $\{ q_i \}$ and $\{ p_i \}$,
\begin{equation}
D(q||p) \equiv \sum_i q_i \log \frac{q_i}{p_i} \geq 0, 
\end{equation} 
assuring the second inequality to hold.
The equality of the first inequality in Eq.~(\ref{eq:eff}) is met for the case of perfect coupling, where the chemical potential of coupled reaction contributes 100\% in altering the balance in the system.
This is a realizable situation in real systems as opposed to the slow and deterministic operation limit $\sigma_f \rightarrow 1$.
Thus, Eq.~(\ref{eq:eff}) states the existence of a reachable bound in autonomous free energy transduction efficiency, which is more restrictive than the limit of the second law.

To derive Eq.~(\ref{eq:eff}), we first examine the free energy gain of the system, $\Delta F$.
Given $\{ U_i \}$, the Gibbs free energy difference of the system's probability distribution to be found as $\{ q_i \}$ instead of $\{ p_i \}$
can be quantified as~\cite{mezard}, 
\begin{equation}
\beta \Delta F = \beta \sum_i (q_i-p_i)U_i - [S(q)-S(p)] = D(q||p), \label{eq:freeene}
\end{equation}
where the Shannon entropy is written as $S(p)= -\sum p_i\log p_i$. Here because the relative entropy is non-negative (0 if and only if $q_i=p_i$ for all $i$),  the equilibrium distribution $\{ p_i \}$ gives the free energy minimum. 
Relative entropy measures the distinguishability of the two probability functions (Sanov's theorem~\cite{kawai,mezard}), showing how unusual it is to find the distribution $\{ q_i \}$ when $\{ p_i \}$ is expected.

\begin{figure}
\begin{center}
\includegraphics[width=85mm]{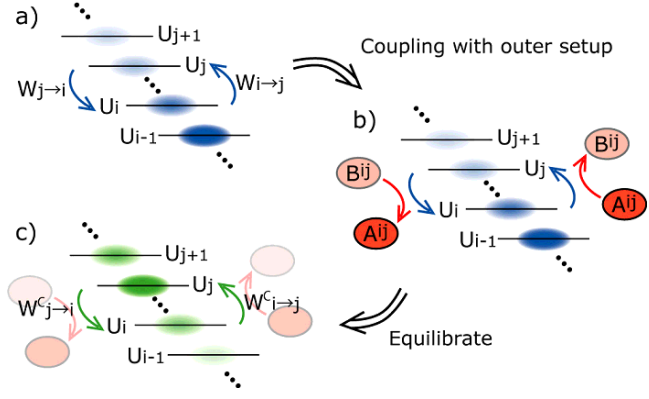}
\caption{\label{fig:balance}(color online). Scheme of revising the distribution of the system with energy states $\{ U_i \}$.
(a) The original detail balanced transitions $w_{i \rightarrow j}$ (blue arrows, Eq.~(\ref{eq:oribalance})), and the equilibrium distribution $\{ p_i \}$ (blue clouds).
(b) Chemical reactions (red, Eq.~(\ref{eq:molecule})) are coupled to the original detail balanced transitions to effectively change the transition rates.
(c) Original distribution is altered and equilibrates to a new distribution $\{q_i \}$ (green clouds), because of new transition rates $w^c_{i \rightarrow j}$ (green arrows, Eq.~(\ref{eq:newbalance})).
For the system this will be a free energy gain of amount $D(q||p)$ (Eq.~(\ref{eq:freeene})) 
and for the outer setup (containing the coupled reactions) 
the loss of at least $D(q||p)+D(p||q)$ (Eq.~(\ref{eq:freeeneg})).}
\end{center}
\end{figure}

Next we derive the free energy  cost, $\Delta F_c$, consumed in the outer setup, which is required for the system to obtain such free energy gain. 
The outer setup that links to the system consists of chemical
species $\{{\rm A}^{ij}$, ${\rm B}^{ij}\}$ with concentrations $\{[{\rm A}^{ij}]$, $[{\rm B}^{ij}]\}$. Chemical
reactions coupled to the transitions between the states in the
system are defined as (see fig.~\ref{fig:balance})
\begin{equation}
{\rm state}\ i + {\rm A}^{ij} \overset{w^c_{i\rightarrow j}}{\underset{w^c_{j\rightarrow i}}{\rightleftharpoons}}  {\rm state}\ j + {\rm B}^{ij}. \label{eq:reac}
\end{equation}
$w^c_{i\rightarrow j}$ are the new Markovian transition rates between energy states $(i, j)$, 
including the effect of the coupled chemical reactions.
The chemical potential (i.e., Gibbs free energy) change $\Delta \mu_{ij}$,
\begin{eqnarray}
{\rm A}^{ij} \longrightarrow {\rm B}^{ij}-\Delta \mu_{ij}, \ \ {\rm B}^{ij} \longrightarrow {\rm A}^{ij}-\Delta \mu_{ji}, \label{eq:molecule}
\end{eqnarray}
in the dilute limit is
 $\Delta \mu_{ij}= -\Delta \mu_{ji}= \Delta \mu_{ij}^0+ \beta^{-1} \log \frac{[{\rm A}^{ij}]}{[{\rm B}^{ij}]}$ where $\Delta \mu_{ij}^0$ is the standard free energy difference of ${\rm A}^{ij}$ and ${\rm B}^{ij}$.
Note that we are considering the temperature $\beta^{-1}$ as being constant throughout.
We assume that there are a sufficient number of molecules  $\{{\rm A}^{ij}, {\rm B}^{ij} \}$
 and that the non coupling reactions ${\rm A}^{ij} \rightleftharpoons {\rm B}^{ij}$ are slow enough, 
$\Delta \mu_{ij}$ 
does not change throughout the time period of interest;
in other words, $\{[{\rm A}^{ij}]$, $[{\rm B}^{ij}]\}$ is fixed.
Thus, our outer setup containing molecules $\{ {\rm A}^{ij}, {\rm B}^{ij} \}$ serves as the chemical potential bath.
In reality, the number of molecules in the reactions may not be conserved, as in the case of ATP $\rightleftharpoons$ ADP+Pi.
However this does not affect the free energy cost, and therefore we may think of reactions (\ref{eq:reac}) without the loss of generality.

Coupling reactions with this outer setup alter the detailed balance condition of the system.
Considering the initial equilibrium setup, detailed balance between the original Markovian transition rates
of the system $w_{i \rightarrow j}$, should satisfy
\begin{equation}
\frac{w_{i\rightarrow j}}{w_{j\rightarrow i}} = e^{\beta[U_i-U_j]}. \label{eq:oribalance}
\end{equation}
New transition rates in the system after being coupled to the
outer setup reactions $w^c_{i \rightarrow j}$ will be altered from Eq.~(\ref{eq:oribalance}), and
will satisfy the relations described using the effective coupling chemical
potentials $\tilde{\mu}_{ij}$,
\begin{equation}
\frac{w^c_{i\rightarrow j}}{w^c_{j\rightarrow i}} = e^{\beta[U_i-U_j+\tilde{\mu}_{ij}]}. \label{eq:newbalance}
\end{equation}
The effective couplings $\tilde{\mu}_{ij}$ for attaining the new distribution $\{ q_i \}$
under the conditions of the original distribution being $\{ p_i \}$ and the new balance condition being Eq.~(\ref{eq:newbalance}) are determined by
\begin{equation}
\beta \tilde{\mu}_{ij} = \log \frac{q_i}{q_j} \frac{p_j}{p_i}. \label{eq:muij}
\end{equation}
Note that $\tilde{\mu}_{ij}=\Delta \mu_{ij}$ does not necessary hold, but in general $0 \le \tilde{\mu}_{ij}/\Delta \mu_{ij} \le 1$.
As will be discussed in the last part of this section, in some contexts efficiency is defined as this coupling perfection $\tilde{\mu}_{ij}/ \Delta \mu_{ij}$, which is related to-but different from-the efficiency of free energy transduction (\ref{eq:eff}). 

Before we derive $\Delta F_c$, a few remarks will elucidate our setup of coupling reactions
 at this point.
In general, uncoupled original transitions of rate $w_{i\rightarrow j}$ may coexist with coupled reactions, but 
here
we are interested in the case $w^c_{i\rightarrow j}\gg w_{i\rightarrow j}$ where coupled reactions 
are much more likely to occur than the original transitions.

Uncoupled transitions may occur, but these will only lead to additional consumption of chemical potential, thus reducing the efficiency (without affecting our final result concerning efficiency bounds). We also assume that the steady state of the new dynamics drives no current for the time scale of interest because our consideration is an equilibrium-to-equilibrium transition.
The no-current condition may be satisfied, for example, when any closed path of fast (dominant) reactions
 $i \rightarrow j \rightarrow \cdots \rightarrow k \rightarrow i$ gives the sum of potential changes $\tilde{\mu}_{ij}+ \cdots + \tilde{\mu}_{ki}=0$. This is somewhat artificial restriction for an $N$ state model; however, in the two-state case $N=2$, where there is no restriction since it is always $\tilde{\mu}_{12} + \tilde{\mu}_{21} =0$.

With these assumptions we quantify the overall free energy consumption in the outer setup,
\begin{eqnarray}
\Delta F_c = \sum_{1\leq i<j \leq N} n_{ij}  \Delta \mu_{ij} . 
\end{eqnarray}
Here, $n_{ij}$ represents the
 net transitions between the system's states $i$ and $j$ (coupled to ${\rm A}^{ij}\rightarrow {\rm B}^{ij}$ in the outer setup) that occurred up to equilibration.
Given ${U_i}$ and $\tilde{\mu}_{ij}$ will not determine the details of the Markovian dynamics of equilibration, including  $n_{ij}$,
 because only the ratios between $w^c_{i\rightarrow j}$ and $w^c_{j\rightarrow i}$ are known and not their absolute values.
However, $\Delta F_c$ will be bounded from below by $\sum_{i<j } n_{ij}  \tilde{\mu}_{ij}$ because each term in the sum is positive and  $0\leq \tilde{\mu}_{ij}/ \Delta \mu_{ij} \leq 1$. 
According to our setup, $n_{ij}$ should satisfy $\sum_i n_{ij} = q_j - p_j $ and $n_{ij} = -n_{ji}$.
Thus, this lower bound may be expressed as a symmetrized Kullback-Leibler distance with no dependence on the detail of $n_{ij}$,
\begin{eqnarray}
\beta \Delta F_c &\geq& \beta \sum_{1\leq i<j \leq N} n_{ij}  \tilde{\mu}_{ij}= \frac{1}{2}\sum_{1\leq i, j \leq N} n_{ij}  \log \frac{q_i}{q_j} \frac{p_j}{p_i} \nonumber \\
&=& \sum_{i} (q_i-p_i)\log \frac{q_i}{p_i} = D(q||p)+D(p||q). \label{eq:freeeneg}
\end{eqnarray}
Equality holds when $\tilde{\mu}_{ij} = \Delta \mu_{ij}$ for all $(i,j)$, which is the perfect coupling case. Free energy gain Eq.~(\ref{eq:freeene}) and the minimal cost Eq.~(\ref{eq:freeeneg}) gives our result of the efficiency bound Eq.~(\ref{eq:eff}).

The physical reason for the appearance of the symmetrized Kullback-Leibler distance may be explained  as follows.
The second term in Eq.~(\ref{eq:freeeneg}), $D(p||q)\geq0$, is the extra free energy consumption, because Eq.~(\ref{eq:freeene})
expresses the free energy gain in the system by the transition $\{ p_i \} \rightarrow \{q_i \}$. 
As observed from the {new equilibrium $\{ q_i \}$, the distribution being $\{ p_i \}$  holds a higher free energy of $\beta^{-1} D(p||q)$, because this is the opposite of Eq.~(\ref{eq:freeene}). From this point of view, $\{ p_i \} \rightarrow \{ q_i \}$ is merely a relaxation process; hence, the additional term in Eq.~(\ref{eq:freeeneg}) appears as a result of  unutilized and dissipated extra free energy. 
This means that to transduce $\beta^{-1}D(q||p)$ of free energy from the non-equilibrium chemical bath (i.e., the outer setup) to the system by the coupling scheme, there is an accompanying dissipation of at least $\beta^{-1}D(p||q)$ for fulfilling the free-fall picture $\{ p_i \}\rightarrow  \{ q_i \}$.
 Dissipating energy should be supplied by the free energy of the chemical bath; thus, we have the additional symmetrical term appearing in Eq.~(\ref{eq:freeeneg}). 
Note that for the case $D(p||q)/D(q||p) \rightarrow 0$, 
the equality of the second inequality in Eq.~(\ref{eq:eff}) will be met, 
however this occurs only in a specific situation of
an infinitely large free energy transduction, 
as will be seen in the next section for the $N=2$ case.

In theory, efficiency $\sigma_f$ could reach 1 in the quasi-static regime of operational thermodynamics. 
If we hypothetically consider the external chemical potential as a tunable parameter and slowly vary it from 0 to $\tilde{\mu}_{ij}$,
the free energy cost in the outer setup will be
\begin{eqnarray}
\sum _i \int ^{q_i} _{p_i} dx \log \frac{x}{p_i} &=& \sum _i \left[ q_i \log \frac{q_i}{p_i} + p_i - q_i \right] \nonumber \\
= D(q||p) &=& \beta \Delta F, 
\end{eqnarray}
where the extra cost is eliminated because of the quasi-static operation and thus $\sigma_f =1$. 
In the chemical coupling scheme, however, the inevitable dissipation $D(p||q)$ accounts for the 
spontaneous progress of the reaction.

\begin{figure}[b!]
\begin{center}
\includegraphics[width=85mm]{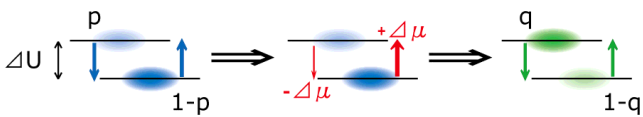}
\caption{\label{fig:alter} (color online). Two-state case of free energy transduction (see Eq.~(\ref{eq:twostate})). Distribution $p:1-p$ is changed to $q:1-q$ by the coupled chemical potential $\Delta \mu$.
.}
\end{center}
\end{figure}

We emphasize that the equality of the first inequality in Eq.~(\ref{eq:eff}) is an obtainable bound in real autonomous systems,
in contrast with the second law bound that requires an operational scheme before it can be reached.
For instance, when the stall force of processive molecular motors are compared to ATP hydrolysis free energy,~\cite{howard} one might think of this efficiency, because in such cases, the new balance Eq.~(\ref{eq:newbalance}) is relative to the free energy consumption $\Delta \mu$ of a single molecule. High coupling efficiency, corresponding to $ \tilde{\mu}_{ij}/ \Delta \mu_{ij}  \sim 1$, has been observed in experiments on molecular motors such as the F1-ATPase~\cite{yasuda}, and perfect coupling is thought to be possible in maximally efficient biological reactions. However, it is important to mention that perfect coupling leads only to the equality of the first inequality in Eq.~(\ref{eq:eff}), and does not guarantee that $\sigma_f =1$. In other words, perfect coupling is the most efficient way of transduction for autonomous systems where optimal controls  such as quasi-static manipulations do not exist.

\textit{Two-State Model.}---Here we investigate the quantities given in the last section, for the $N=2$ case.
In fact, most biological reactions are basically two-state models with a chemical potential bath. Typical examples include molecules transitioning to each other as in high-energy phosphate reactions, conformational states in proteins as in molecular motors/ion pumps and various enzymes, or chemical potential differences across a membrane~\cite{astumian1,astumian2,prost,para}.
\begin{figure}
\begin{center}
\includegraphics[width=55mm]{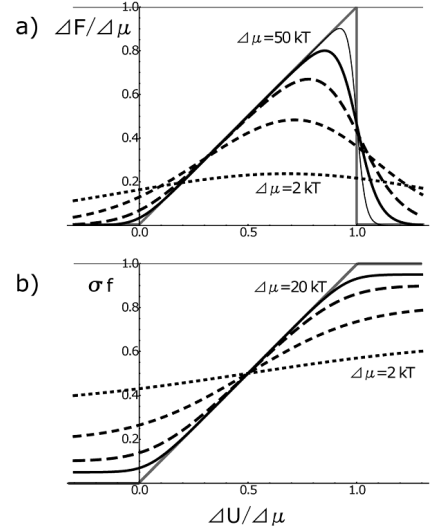}
\caption{\label{fig:freeenefg} Transduced free energy and the efficiency according to the design of the two-state model ($\Delta U$ of fig.~\ref{fig:alter}). (a) Gained free energy in the system with fixed chemical potential $\Delta \mu=50, 20, 10, 5, 2 \ k_B T$, on the order of the height of the peak. (b) Efficiency of free energy transduction (\ref{eq:eff}) for $\Delta \mu=20, 10, 5, 2 \ k_B T$. In both figures the gray line corresponds to the limit $\beta \Delta \mu\rightarrow \infty$.}
\end{center}
\end{figure}

Let the system initially be at probabilities $p:1-p$ for the state-1 and state-2 with the (free) energy difference between the states being $\beta \Delta U = \log\frac {1-p}{p}$ (see fig.~\ref{fig:alter}). 
For the system to reach a new equilibrium with probabilities $q:1-q$, the chemical potential change in the coupling reaction must be a certain amount $-\tilde{\mu}_{12} =\tilde{\mu}_{21} =\Delta \mu$, considering the perfect coupling case.
An average number $-n_{12}= n_{21}=q-p$ of molecules will be consumed in the chemical potential bath; thus,
\begin{eqnarray}
\beta \Delta F &=& D(q||p) = q\log \frac{q}{p} + (1-q)\log \frac{1-q}{1-p}, \nonumber \\
\beta \Delta \mu &=& \log \frac{q(1-p)}{p(1-q)},\nonumber \\
\beta \Delta F_c &=& D(q||p)+D(p||q)=(q-p) \Delta \mu . \label{eq:twostate}
\end{eqnarray}
For a fixed $\Delta \mu$, received ($\Delta F$) and consumed ($\Delta F_c$) free energies, change according to the
initial setting of the system ($\Delta U$ in fig.~\ref{fig:alter}).
In fig.~\ref{fig:freeenefg}, we plot $\Delta F$ and the free energy transition efficiency $\sigma_f$ as a function of $\Delta U$.
}Assuming $\Delta \mu>0$ without the loss of generality,
$\sigma_f$ is a monotonically increasing function of $\Delta U$ approaching
\begin{equation}
\sigma_f \rightarrow \overline{\sigma_f} \equiv 1 - \frac{1}{x}+\frac{1}{e^{x} -1} \  \leq 1, \label{eq:effmax}
\end{equation}
where $x=\beta \Delta \mu$ and for $\Delta U \rightarrow \infty$ giving the upper bound of $\sigma_f$ (fig.~\ref{fig:function}).
Only in the case of $\Delta \mu \rightarrow \infty$ does $\sigma_f$ reach 1, because $\sigma_f \leq \overline{\sigma_f}$ for all $\Delta U$.
In fact, the condition for $\sigma_f = 1$ is $\Delta U/\Delta \mu \geq 1$ with $\beta \Delta \mu \rightarrow \infty $ 
(see the gray plot in fig.~\ref{fig:freeenefg} (b)).
\begin{figure}
\begin{center}
\includegraphics[width=50mm]{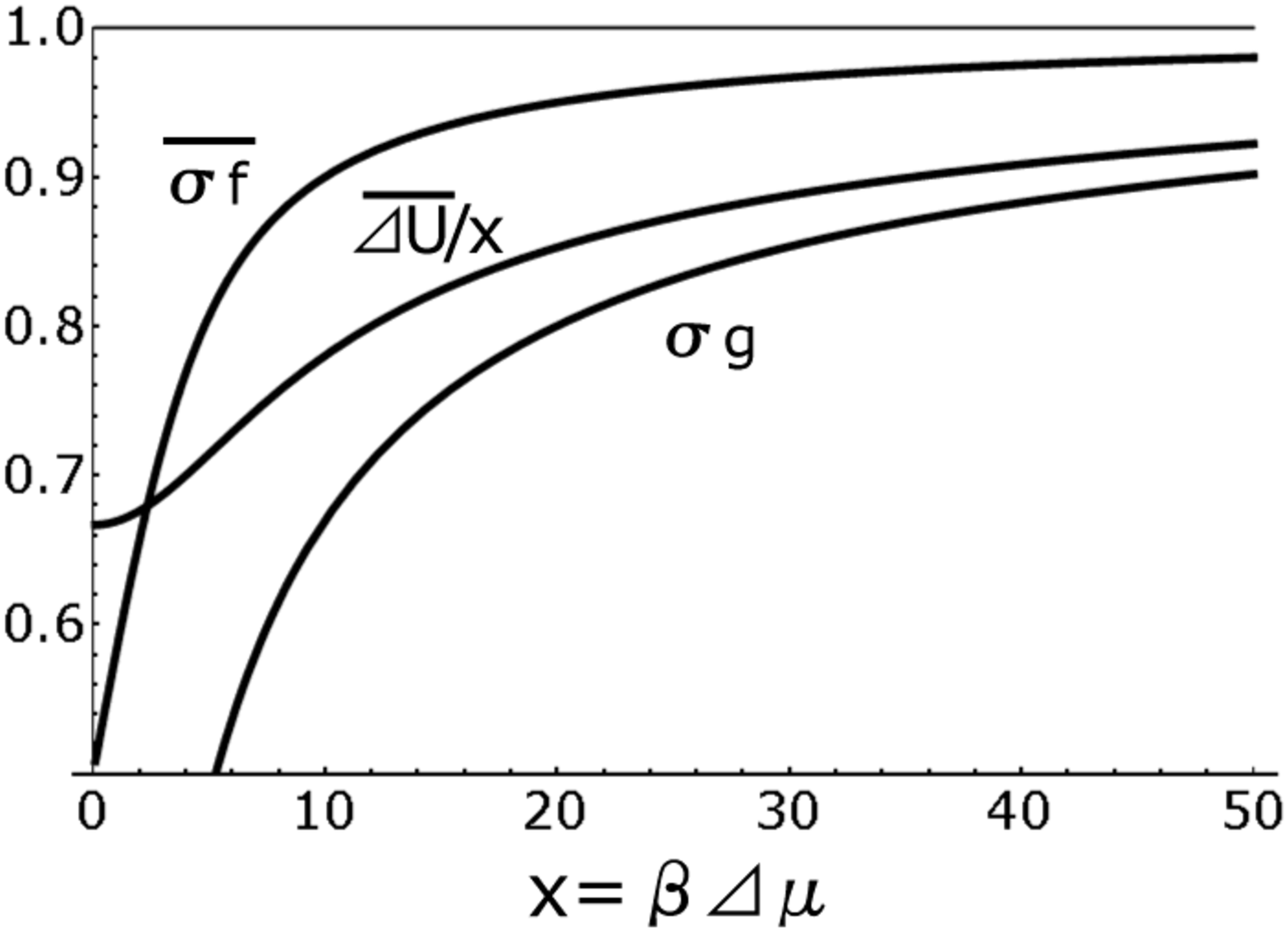}
\caption{\label{fig:function}Efficiencies (\ref{eq:effmax}), (\ref{eq:tramax}) and the optimal $\Delta U$ (\ref{eq:opt}). $\sigma_g$ and $\overline{\Delta U}/x$ are the ordinate and abscissa, respectively, of the peaks in  fig.~\ref{fig:freeenefg} (a), and $\overline{\sigma_f}$ is the $\Delta U/\Delta \mu \rightarrow \infty $ limit in fig.~\ref{fig:freeenefg} (b)}
\end{center}
\end{figure}

Another efficiency of interest is the maximal free energy that could be transduced out of the given chemical potential,
\begin{equation}
\sigma_g \equiv \frac{\max_{\Delta U} \Delta F}{|\Delta \mu|}, 
\end{equation}
which is the peak value of each of the plots in the fig.~\ref{fig:freeenefg} (a).
Because $\Delta F_c<|\Delta \mu|$, we see that $\sigma_g < \sigma_f$.
The explicit expression as a function of $x=\beta \Delta \mu$ can then be derived;
\begin{equation}
\sigma_g (x) = \frac{1}{1-e^{-x}} + \frac{1}{x} \left[ \log \frac{1-e^{-x}}{x}-1 \right], \label{eq:tramax}
\end{equation}
where the maximum of $\Delta F$ is given at
\begin{equation}
\beta \Delta U = \overline{\Delta U}(x) \equiv x - \log \frac{x e^{x} +1 -e^{x}}{e^{x} -1 -x}. \label{eq:opt}
\end{equation}

From these expressions, both functions $\sigma_g (x)$ and $\overline{\Delta U}(x)/x$ are found to be smaller than 1, tending to $1-\log x/x$ for large enough $x$, as plotted in fig.~\ref{fig:function}.
Consequently, we see $\sigma_g$ reaching as low as $ \sim 0.8$ in the case of $\Delta \mu \sim 20 k_B T$,
which is thought to be the ATP hydrolysis free energy in a typical intracellular culture,
 and less than 0.9 even in the case of  $\Delta \mu \sim 50 k_B T$, which is the maximum energy scale of single molecular biochemical reactions (fig.~\ref{fig:function}).
This shows the inefficiency 
of transmitting finite free energy from the outer setup to the system by a non-operative procedure.
The expression of $\overline{\Delta U}(x)$ provides an idea for the design of an ideal free energy transduction: 
the optimal two-molecular state system for transducing a free energy $\Delta \mu$ is a system with a standard free energy difference $\overline{\Delta U}$.
This design principle may be adopted 
in biological systems, 
for example, in the metabolic pathway where free energy may not be wasted in successive transductions
or in the signal transduction pathway where information must be transmitted with high efficiency.

In summary, we derived the explicit formula of the free energy transduction efficiency limit as Eq.~(\ref{eq:eff}) in a chemical coupling regime. 
The symmetrized Kullback-Leibler distance in Eq.~(\ref{eq:freeeneg}) lowers the bound of efficiency by a non-zero counter term, which appears as a consequence of the procedure of free energy transduction being autonomous.
For the two-state model, derived expressions of maximal efficiencies were shown to remain low for finite free energy cases,
and the situation of optimal transduction was pointed out for its promise for research in real biological systems.

We believe the efficiency bound (\ref{eq:eff}) could be derived for various setups 
as in the case of non-equilibrium work relations~\cite{jar,trep,crooks,hatano}, 
because the reason for the symmetrized Kullback-Leibler distance to enter the formula seems sufficiently general.

Our view shows the importance of distinguishing between the ultimate bound of the second law,
which can be achieved only in the macroscopically/quasi-statically operating case,
and the realistic tighter bound imposed on autonomous systems.
Because the loss of free energy in the form of $\Delta F_c - \Delta F \geq \beta^{-1} D(p||q)$ will accumulate in the case of successive transduction,
it should be interesting to learn how biochemical networks maximize its efficiency of 
information/free energy transmission in their series of reactions.

\begin{acknowledgments}
We would like to thank Takahiro Sagawa for many fruitful discussions and reading of the manuscript.
This work was supported by a Grant-in-Aid for Scientific Research (No. 21244061)
 from JSPS.
\end{acknowledgments}

\end{document}